\documentclass[a4paper,accepted=2024-06-25]{quantumarticle}
\pdfoutput=1
\usepackage[utf8]{inputenc}
\usepackage[english]{babel}
\usepackage[T1]{fontenc}
\usepackage{amsmath}
\usepackage{amssymb}
\usepackage{physics}
\usepackage{hyperref}
\usepackage{caption}
\usepackage{graphicx}
\usepackage{ifthen}
\graphicspath{{figures/}}
\DeclareMathOperator{\Var}{Var} % Variance
\DeclareMathOperator{\Cov}{Cov} % Covariance
\DeclareMathOperator*{\argmin}{arg\,min}

\newcommand{\secref}[1]{\hyperref[#1]{Section~\ref{#1}}}
\newcommand{\appref}[1]{\hyperref[#1]{Appendix~\ref{#1}}}
\newcommand{\tabref}[1]{\hyperref[#1]{TABLE~\ref{#1}}}
\newcommand{\figref}[2][]{\hyperref[#2]{FIG.~\ref{#2}#1}}
\newcommand{\codeavailability}[1]{%
	\section*{Code Availability}%
	#1%
}

\begin{document}

%
% ****** Authors ******
\author{Pontus Vikstål}
\email[e-mail:~]{vikstal@chalmers.se}
\affiliation{Wallenberg Centre for Quantum Technology, Department of Microtechnology and Nanoscience, Chalmers University of Technology, 412 96 Gothenburg, Sweden}

\author{Giulia Ferrini}
\affiliation{Wallenberg Centre for Quantum Technology, Department of Microtechnology and Nanoscience, Chalmers University of Technology, 412 96 Gothenburg, Sweden}

\author{Shruti Puri}
\affiliation{Department of Applied Physics, Yale University, New Haven, Connecticut 06511, USA}
\affiliation{Yale Quantum Institute, Yale University, New Haven, Connecticut 06511, USA}

%
% ****** Title ******
\title{Study of noise in virtual distillation circuits for quantum error mitigation}

%
% ****** Date ******
%\date{28-10-2022}

%
% ****** Abstract ******
\begin{abstract}
    Virtual distillation has been proposed as an error mitigation protocol for estimating the expectation values of observables in quantum algorithms. It proceeds by creating a cyclic permutation of $M$ noisy copies of a quantum state using a sequence of controlled-swap gates. If the noise does not shift the dominant eigenvector of the density operator away from the ideal state, then the error in expectation-value estimation can be exponentially reduced with $M$. In practice, subsequent error mitigation techniques are required to suppress the effect of noise in the cyclic permutation circuit itself, leading to increased experimental complexity. Here, we perform a careful analysis of the effect of uncorrelated, identical noise in the cyclic permutation circuit and find that the estimation of expectation value of observables are robust against dephasing noise. We support the analytical result with numerical simulations and find that $67\%$ of errors are reduced for $M=2$, with physical dephasing error probabilities as high as $10\%$. Our results imply that a broad class of quantum algorithms can be implemented with higher accuracy in the near-term with qubit platforms where non-dephasing errors are suppressed, such as superconducting bosonic qubits and Rydberg atoms.
\end{abstract}

%
% ****** Make title ******
\maketitle

%=============================================
% INTRODUCTION
%=============================================
\section{\label{sec:introduction}Introduction}
Fault-tolerant quantum error correction is necessary for scalable quantum computation~\cite{lidar_quantum_2013}, however the associated hardware-performance requirements and resource overheads are hard to meet with the noisy intermediate-scale quantum processors available today. Consequently, for near-term applications alternative techniques to mitigate the effect of noise have been developed. Some of these techniques are based on scaling noise~\cite{li_efficient_2017,temme_error_2017,endo2018practical,kandala2019error} or learning about the effect of noise to predict the noise-free behavior of the quantum protocol~\cite{strikis2021learning,czarnik2021error}, while others exploit the symmetry properties of the noise-free quantum circuit to flag errors~\cite{bonet2018low,mcardle2019error,sagastizabal2019experimental,google2020hartree,huggins2021efficient}. Algorithm- and noise-specific error mitigation techniques have also been proposed~\cite{o2021error,maciejewski2020mitigation}. 
\par
Recently an error mitigation scheme known as virtual distillation, or error suppression by derangement, has been shown to achieve an exponential suppression of errors in the estimation of the expectation value of an observable~\cite{cotler_quantum_2019,koczor_exponential_2021,huggins_virtual_2021}. The key idea behind this protocol is to compute the expectation value of an observable by performing measurements on a cyclic-permutation of $M$ copies of a noisy quantum state. If the effect of noise is to mix the ideal noise-free state with orthogonal error states, then symmetries of the cyclic-permutation state suppress the contribution to the expectation value from the error states exponentially in $M$.
\par
The most straightforward approach to virtual distillation is to prepare the cyclic-permutation state using an auxiliary qubit and controlled-SWAP (CSWAP) gates. In practice, this circuit will be prone to errors, limiting the accuracy of expectation-value estimation without resorting to further noise mitigation techniques, like zero-noise extrapolation ~\cite{li_efficient_2017,temme_error_2017, koczor_exponential_2021}. However, zero-noise extrapolation not only adds to the sampling cost, but also considerably increases the circuit complexity as it requires the ability to scale the noise strength in the quantum circuit either by scaling gate times or by adding more gates into the circuit~\cite{endo2018practical,kandala2019error,dumitrescu_cloud_2018,otten_recovering_2019,giurgica-tiron_digital_2020}. Thus, in this paper we further investigate the effect of noise in the virtual distillation circuit and determine analytically conditions under which its faults may be less detrimental, obviating the need for additional error mitigation. We corroborate our findings with numerical simulations of the Quantum Approximate Optimization Algorithm (QAOA). Noise in virtual distillation circuits was previously considered numerically in the context of Heisenberg quench~\cite{huggins_virtual_2021} as well as for the variational quantum eigensolver~\cite{jnane_multicore_2022}, displaying robustness of the error mitigation procedure.
\par
We consider three commonly studied types of noise: depolarizing, dephasing, and amplitude damping noise and find that the mitigated expectation value with virtual distillation is robust against dephasing noise for an arbitrary even number of copies $M$. We support our analysis with numerical simulation of QAOA for solving a MaxCut problem of partitioning the set of vertices in a given graph into two subsets such that the number of edges shared between the two partitions is maximized, for the case of two copies, $M = 2$. QAOA is implemented by preparing a variational quantum state using a short-depth quantum circuit and estimating its energy, i.e. the expectation value of the Ising-Hamiltonian associated with the MaxCut problem. The parameters in the circuit are varied until a minimum in the energy landscape is found. In order to overcome the adverse effects of noise in finding a state that minimizes the energy, we combine the QAOA protocol with virtual distillation. We find that the error in estimating the energy with virtual-distillation with $M=2$ is reduced by $67\%$ when the underlying source of noise is single-qubit pure dephasing errors at rate of $10\%$, compared to $20\%$ when the underlying source of noise is single-qubit depolarizing errors at the same rate. Additionally, we found that amplitude damping errors was detrimental to the virtual distillation circuit, resulting in no error mitigation. Our findings imply that virtual distillation in a system in which non-dephasing errors are suppressed compared to dephasing errors is successful at reducing errors in expectation-value estimation of observables diagonal in the computational basis without additional error mitigation schemes. It is known that such an error channel is relevant for Kerr-cat qubits in superconducting microwave circuits~\cite{puri_bias_2020,grimm2020stabilization} and Rydberg atomic qubits~\cite{iris_hardware_2022} not only when the qubits are idle but also during implementation of Toffoli and controlled-not gates. These two gates can be combined to implement a CSWAP~\cite{smolin_1996_five} and thus it is possible to realize robust virtual distillation in these platforms. 
\par
This paper is organized as follows: In \secref{sec:virtualdistillation} we review the virtual distillation protocol. We analyze the effect of noise in the virtual distillation circuit on the estimated expectation value in \secref{sec:noiseinvd}. We support our analysis with numerical simulations in \secref{sec:results} and finally, we give our concluding remarks in \secref{sec:conclusion}.
%=============================================
% VIRTUAL DISTILLATION
%=============================================
\section{\label{sec:virtualdistillation}Virtual Distillation}
We begin this section by establishing the notation used throughout this paper, which is based on Ref.~\cite{huggins_virtual_2021}. A boldfaced superscript, for example $O^{\mathbf{i}}$, will be used to indicate that the operator $O$ acts on the $i^\mathrm{th}$ subsystem. We use superscript with parentheses to indicate an operator acting on multiple subsystems. For instance, $S^{(M)}$ indicates that $S$ acts on $M$ subsystems.
\par
Consider the output density operator, $\rho$, of an $N$-qubit noisy quantum circuit with the spectral decomposition 
\begin{equation}
    \label{eq:spectral}
    \rho = \sum_{k=1}^{d} \lambda_k \dyad{\psi_k}.
\end{equation}
Here $d=2^N$ and $\lambda_k$ is the probability that the system is found in the state $\ket{\psi_k}$ when measuring in the eigenbasis of $\rho$. We assume, for convenience, that the probabilities $\lambda_k$ are listed in descending order $\lambda_1>\lambda_2\ldots>\lambda_d$. In the virtual distillation protocol, raising $\rho$ to the power of $M$ and normalizing it results in a density operator that approaches the dominant eigenvector $\dyad{\psi_1}$ exponentially fast with $M$, i.e.
\begin{equation}
    \label{eq:exponentiation}
    \tilde{\rho}=\frac{\rho^M}{\Tr(\rho^M)} = \frac{\sum_{k=1}^{d} \lambda_k^M \dyad{\psi_k}}{\sum_{k=1}^{d} \lambda_k^M}.
\end{equation}
In virtual distillation, the expectation value of an observable $O$ is estimated with respect to the exponentiated density matrix $\tilde{\rho}$, 
\begin{equation}
    \label{eq:mitigated}
    \expval{O}_\mathrm{mitigated}:={\Tr(O\tilde{\rho})}=\frac{\Tr(O\rho^M)}{\Tr(\rho^M)}.
\end{equation}
When $\ket{\psi_1}$ corresponds to the output of the ideal (noise free) quantum circuit, then $\expval{O}_\mathrm{mitigated}$ approaches the ideal expectation value exponentially fast with $M$. This condition is satisfied when noise in the quantum circuit maps the ideal states to states that are orthogonal to it, otherwise the dominant eigenvector will drift and limit the error suppression efficiency~\cite{koczor_exponential_2021,huggins_virtual_2021}. In general, for a multi-qubit state, single-qubit errors can cause drift of the dominant eigenvector. However, in real-world applications, this drift is expected to be small, as also validated by the numerical simulations in this paper. Furthermore, the severity of this drift, or coherent mismatch, is exponentially smaller than the incoherent decay of fidelity~\cite{koczor_dominant_2021}.
\par
Note that, in virtual distillation, $\expval{O}_\mathrm{mitigated}$ is calculated without explicitly preparing the state $\tilde{\rho}$, hence the name ``virtual''. Instead, virtual distillation uses $M$ copies of the state $\rho$ together with collective measurements that only allow symmetric states of the form $\ket{\psi}\otimes\ket{\psi}\ldots\ket{\psi}$ to contribute to the expectation value of $O$. More specifically, in Ref.~\cite{huggins_virtual_2021} it was shown that Eq.~\eqref{eq:mitigated} is equivalent to
\begin{equation}
    \label{eq:estimated}
    \expval{O}_\mathrm{mitigated} := \frac{\Tr(O^{(M)}S^{(M)}\rho^{\otimes M})}{\Tr(S^{(M)}\rho^{\otimes M})},
\end{equation}
where $O^{(M)}$ is the symmetrized version of the operator $O$,
\begin{equation}
    \label{eq:symmeterized}
    O^{(M)} = \frac{1}{M}\sum_{\mathbf{i}=1}^M O^\mathbf{i},
\end{equation}
and $S^{(M)}$ is the cyclic shift operator that act on all $M$ subsystems. Its effect is only to let symmetric states of $\rho^{\otimes M}$ to contribute to the expectation value of Eq.~\eqref{eq:estimated},
\begin{equation}
    \label{eq:cyclicshift}
    S^{(M)} \ket{\psi_1}\otimes \ket{\psi_2}\ldots\ket{\psi_M} 
    = \ket{\psi_2}\otimes \ket{\psi_3}\ldots\ket{\psi_1}.
\end{equation}
To measure the observable $S^{(M)}$ in Eq.~\eqref{eq:estimated} virtual distillation uses a procedure similar to the Hadamard test~\cite{aharonov_polynomial_2009}. The procedure begins by preparing $M$ collective copies of the state $\rho$ together with an auxiliary qubit in the state $\ket{+}=(\ket{0}+\ket{1})/\sqrt{2}$. Next, a sequence of CSWAP gates applies $S^{(M)}$ to the $M$ copies of $\rho$ conditioned on the auxiliary qubit being in state $\ket{1}$. Finally the auxiliary qubit is measured in the $X$-basis and its expectation value equals to $\Tr(S^{(M)}\rho^{\otimes M})$, i.e. the denominator of Eq.~\eqref{eq:estimated}. Since $O^{(M)}$ commutes with $S^{(M)}$, these two operators can be simultaneously diagonalized, allowing them to be measured at the same time. By also measuring $O^{(M)}$ on the $M$ subsystems $\rho$, the measurement outcome can be used together with the measurement outcome from the auxiliary qubit to estimate the numerator $\Tr(O^{(M)}S^{(M)}\rho^{\otimes M})$.
%=============================================
% ERROR MODELS
%=============================================
\section{\label{sec:noiseinvd}Noise in virtual distillation circuits}
We model a noisy gate in the virtual distillation circuit as an ideal gate followed by independent and identical single-qubit errors acting on each qubit participating in the gate. We examine three types of single-qubit noise channels. The first one is the depolarizing channel which describes a process where information is completely lost with some probability $\epsilon$, and is given by~\cite{nielsen_quantum_2010}
\begin{equation}
    \label{eq:depolarizing-channel}
    \Lambda_\mathrm{dep}(\rho) = \qty(1-\epsilon)\rho + \frac{\epsilon}{3}(X\rho X + Y\rho Y + Z\rho Z),
\end{equation}
where $\{X,Y,Z\}$ are the Pauli operators and $\epsilon$ is the error probability. The second one is the pure-dephasing channel which is a biased noise channel\footnote{Of course we could have chosen an error channel with biased X- or Y-noise but we can always redefine the computational basis states on Bloch sphere and call all of these Z-biased noise.}and describes loss of phase information with a probability $\epsilon$,
\begin{equation}
    \label{eq:dephasing-channel}
    \Lambda_\mathrm{Z}(\rho) = \qty(1-\epsilon)\rho + \epsilon Z\rho Z.
\end{equation}
The third and final channel that we consider is the amplitude damping channel which is characterized by energy dissipation to the ground state over time. Although no analytical expression of the mitigated expectation value for the amplitude damping channel is derived in this work, it is defined as follows:
\begin{equation}
    \label{eq:amplitude-damping-channel}
    \Lambda_\mathrm{amp}(\rho) = K_1\rho K_1^\dagger + K_2\rho K_2^\dagger.
\end{equation}
where the Kraus operators $K_1$ and $K_2$ are given by
\begin{equation}
    K_1 = 
    \begin{pmatrix}
        1 & 0 \\
        0 & \sqrt{1-\gamma}
    \end{pmatrix},
    \quad
    K_2 =
    \begin{pmatrix}
        0 & \sqrt{\gamma} \\
        0 & 0
    \end{pmatrix}.
\end{equation}
with
\begin{equation}
    \gamma \equiv 4(\sqrt{1-\epsilon}+\epsilon-1).
\end{equation}
We use these definitions of the error channels because their average channel fidelities are the same for a given $\epsilon$, allowing for a consistent comparison across the different noise models.

In the next section we will present analytical results on how the depolarizing and dephasing noise channels affect the mitigated expectation value of virtual distillation as well as their associated variances. For the amplitude damping channel, instead, corresponding analytical expressions could not be obtained, and numerical results on that channel will be presented in the later \mbox{\secref{sec:results}}.
%================BEGIN FIGURE=================
\begin{figure*}
    \centering
    \includegraphics[]{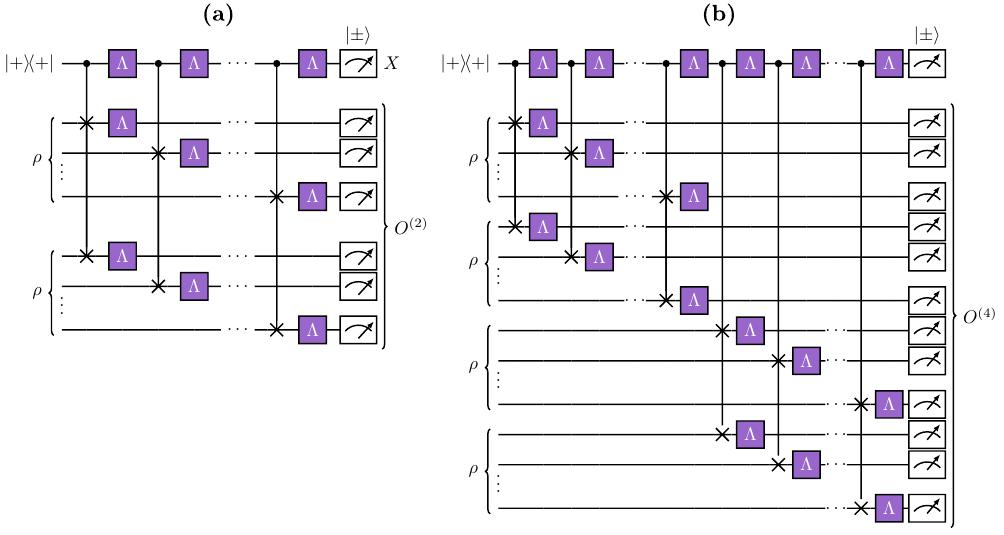}
    \caption{The virtual distillation circuit for \textbf{(a)} $M=2$ and \textbf{(b)} $M=4$,  with single-qubit errors $\Lambda$.}
    \label{fig:vd}
\end{figure*}
%=================END FIGURE==================
%=============================================
% MITIGATED EXPECTATION VALUES
%=============================================
\subsection{Noisy mitigated expectation values}
In this section we provide the main results of this paper. We derive an expression for the noisy mitigated expectation value for even number of copies. For any number of copies $M$, the cyclic shift operator $S^{(M)}$ factorizes into a tensor product of $MN/2$ number of SWAPs, and its controlled version factorizes into a product of $MN/2$ CSWAP-gates. For the virtual distillation circuit, we assume that a single-qubit noise channel is applied after each gate to the qubits involved, see \mbox{\figref{fig:vd}}. For even number of copies $M$, only one swap per subsystem is required, as for example shown for the case of $M=4$ in \mbox{\figref[b]{fig:vd}}. As a consequence, for the  case of even number of copies we find the following analytical expression for the mitigated expectation value:
\begin{equation}
    \label{eq:mitigated-expval-even-m}
    \expval{O}^\Lambda_\mathrm{mitigated}
    = \frac{\Tr(\bar\Lambda(O)\rho^M)}{\Tr(\rho^M)},
\end{equation}
where $\bar \Lambda = \Lambda \otimes \ldots \otimes \Lambda$ is a tensor product of $N$ single-qubit error channels and $\Lambda\in\{\Lambda_\mathrm{dep},\Lambda_\mathrm{Z}\}$. The details of the calculations are provided in \mbox{\appref{app:mitigated-expvals}}. For the case of an odd number of copies, the calculation is more involved. Consider for instance the case of three copies. In this case, one of the copies needs to be swapped twice, making the mathematical derivation of the mitigated expectation value significantly more difficult.
From Eq.~\eqref{eq:mitigated-expval-even-m} we see that the influence of errors on the mitigated expectation value will depend on the observable $O$. Since a general observable on $N$ qubits can be expressed as a sum of $N$-qubit Pauli strings from the set $\{I,X,Y,Z\}^{\otimes N}$, it is sufficient to consider $O\in \{I,X,Y,Z\}^{\otimes N}$. In this case we find that the mitigated expectation value for the two types of noise are given by
\begin{align}
    \expval{O}_\mathrm{mitigated}^{\Lambda_\mathrm{dep}}
    &= \qty(1-\frac{4}{3}\epsilon)^k\frac{\Tr(O\rho^M)}{\Tr(\rho^M)}, \\
    \expval{O}_\mathrm{mitigated}^{\Lambda_\mathrm{Z}}
    &= \qty(1-2\epsilon)^{k'}\frac{\Tr(O\rho^M)}{\Tr(\rho^M)},
\end{align}
where $k$ is the number of $\{X,Y,Z\}$ Pauli matrices in the tensor product of $O$, and $k'$ is the number of $\{X,Y\}$ Pauli matrices in the tensor product of $O$. Thus we see that errors in the virtual distillation circuit only attenuate the expectation values. We will study the sampling cost of this attenuation in the next section. The attenuation can, in principle, be overcome by error mitigation techniques like polynomial extrapolation. Importantly, we note that the mitigated expectation value of an observable $O$ that is only a tensor product of Pauli $Z$-operators, so that $k'=0$, is completely immune to pure dephasing in the distillation circuit. This is typically the case for variational algorithms for combinatorial optimization~\mbox{\cite{farhi_qaoa_2014,lucas_ising_2014,vikstal_2020}} or electronic structure calculation of molecular Hamiltonians~\cite{xia_electronic}. Moreover by performing a local Clifford transformations on the state $\rho$ before sending it into the virtual distillation circuit, it is possible to measure any Pauli observable without attenuating the mitigated expectation value. 
%================BEGIN FIGURE=================
\begin{figure*}[t]
    \centering
    \includegraphics{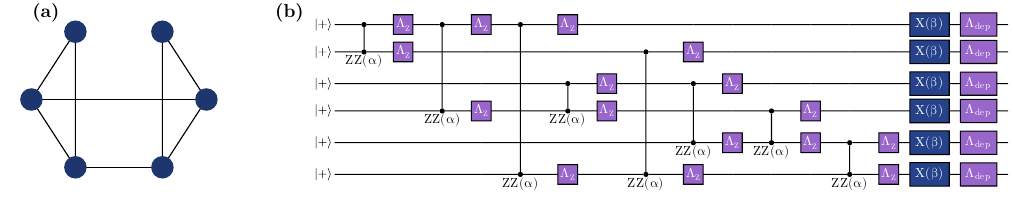}
    \caption{\textbf{(a)} A graph illustrating one instance of the MaxCut problem used in our study. Nodes represent qubits and edges denote interactions between them. \textbf{(b)} The corresponding QAOA circuit for $p=1$ with dephasing noise for the graph in (a). Here $\textsc{ZZ}(\alpha)$ on qubit $j$ and $k$ is $e^{-i\alpha Z_j Z_k / 2}$, and $\textsc{X}(\beta)$ on qubit $j$ is $e^{-i\beta X_j}$. $\Lambda_\mathrm{Z}$ is a single qubit dephasing-channel and $\Lambda_\mathrm{dep}$ is a single-qubit depolarizing channel to account for the fact that $Z$-errors do not commute with the $X$-rotations. The total gate count in the figure is $6$ single-qubit gates and $8$ two-qubit gates, excluding the error-channels.}
    \label{fig:qaoa-circuit}
\end{figure*}
%=================END FIGURE==================
%=============================================
% VARIANCE
%=============================================
\subsection{Variance of the estimator}
We now turn our attention to investigating the sample variance in the estimation of $\Tr(O^{(M)}S^{(M)}\rho^{\otimes M})/\allowbreak\Tr(S^{(M)}\rho^{\otimes M})$. Here we limit ourselves to the case of $M=2$, because the case of general number of copies involves calculating expectation values of the sort $\expval{(O^{(M)})X_\mathrm{aux})^2}$ (where $X_\mathrm{aux}$ is the Pauli-$X$ on the auxiliary qubit), which becomes quickly difficult for $M>2$.

There exists no closed expression of the sample variance of the quotient between two random variables, but an approximated one can be obtained by Taylor-expanding the variance around the mean. If we let $\bar x$ denote the sample mean of $\Tr(O^{(2)}S^{(2)}\rho^{\otimes 2})$, and $\bar y$ the sample mean of $\Tr(S^{(2)}\rho^{\otimes 2})$, then the following unbiased estimator can be constructed
\begin{equation}
    \label{eq:estimator}
    \tilde\theta:=\frac{\bar x}{\bar y}.
\end{equation}
Given sufficiently many samples $R$, the variance of this estimator, for the noiseless case of \figref[a]{fig:vd}, can be approximated as~\cite{huggins_virtual_2021}
\begin{multline}
    \label{eq:mitigated-variance}
    \Var\qty(\mathrm{estim.})\approx \frac{1}{R}
    \Bigg(\frac{\Tr(O^2\rho)}{2\Tr(\rho^2)^2} 
    + \frac{\Tr(O\rho)^2}{2\Tr(\rho^2)^2}
    \\
    + \frac{\Tr(O\rho^2)^2}{\Tr(\rho^2)^4}
    - 2\frac{\Tr(O\rho^2)}{\Tr(\rho^2)^3}\Tr(O\rho)\Bigg).
\end{multline}
Our next step is to find how this variance changes when there is noise in the virtual distillation circuit. Starting with dephasing errors, we find in \appref{app:variance} that the variance of the estimator for a Pauli-string is
\begin{widetext}
\begin{multline}
    \Var_{\Lambda_\mathrm{Z}}\qty(\mathrm{estim.})=
    \frac{1}{R(1-2\epsilon)^{2N}}
    \Bigg(\frac{\Tr(\bar\Lambda_\mathrm{Z}(O^2)\rho)}{2\Tr(\rho^2)^2}
    + \frac{\Tr(\bar\Lambda_\mathrm{Z}(O)\rho)^2}{2\Tr(\rho^2)^2}
    + \frac{\Tr(\bar\Lambda_\mathrm{Z}(O)\rho^2)^2}{\Tr(\rho^2)^4}
    \\- 2\frac{\Tr(\bar\Lambda_\mathrm{Z}(O)\rho^2)}{\Tr(\rho^2)^3}\Tr(\bar\Lambda_\mathrm{Z}(O)\rho)\Bigg),
\end{multline}
\end{widetext}
which is the same as Eq.~\eqref{eq:mitigated-variance} but scaled by a factor $(1-2\epsilon)^{-2N}$ and with $O$ replaced by $\bar\Lambda_\mathrm{Z}(O)$. In the case of $O$ being a tensor product of $Z$-operators, $\bar \Lambda_\mathrm{Z}(O) = O$, the expression reduces to
\begin{equation}
    \label{eq:variance-dephasing}
    \Var_{\Lambda_\mathrm{Z}}\qty(\mathrm{estim.})=\frac{1}{(1-2\epsilon)^{2N}}\Var\qty(\mathrm{estim.}).
\end{equation}
This equation shows that $(1-2\epsilon)^{2N}$ extra circuit repetitions are required for reaching the same level of precision compared to noise free virtual distillation. However, when $\epsilon\ll 1$ the denominator can be expanded as $(1-2\epsilon)^{-2N}\approx 1 + 4N\epsilon+\mathcal{O}(\epsilon^2)$, which shows that $4N\epsilon$ extra circuit repetitions are required for small error probabilities. In the presence of depolarizing errors no simple expression for the variance of the estimator is found, but we note that for a global depolarizing channel acting after each CSWAP gate a similar expression to Eq.~\eqref{eq:variance-dephasing} was found in Ref.~\cite{czarnik_qubit-efficient_2021}.
%=============================================
% RESULTS
%=============================================
\section{\label{sec:results}Numerical Results}
To corroborate our analytical results we will now perform numerical experiments by simulating the quantum approximate optimization algorithm (QAOA), solving 6-qubit MaxCut problems on 30 randomly generated Erdős--Rényi graphs~\cite{erdos_random_1959} with 6-vertices and edge probability of $0.5$ (see \mbox{\figref[a]{fig:qaoa-circuit}}).
\par
The MaxCut problem is defined by a graph $G=(V,E)$, where $V$ is the set of vertices and $E$ is the set of edges. The objective of MaxCut is to partition the set of vertices into two subsets, such that the number of edges from one partition to the other is maximum. The problem can be reformulated as finding the ground state of a Hamiltonian
\begin{equation}
    \label{eq:maxcut}
    H_\mathrm{MaxCut} = -\frac{1}{2}\sum_{i,j\in E}(I-Z_iZ_j),
\end{equation}
where $Z_i$ and $Z_j$ are Pauli $Z$ matrices.
\par
We simulate both QAOA and the virtual distillation circuits in the presence of either single-qubit depolarizing, dephasing or amplitude damping errors. Moreover, in order to also make a comparison that is independent of the state input to the three noisy virtual distillation circuits, we use the same 30 graphs to create a mixed state involving the state corresponding to the maximum cut and a thermal state for each graph. The mixed state is then used as input to the virtual distillation circuit, and we benchmark its performance in the presence of either single-qubit depolarizing, dephasing or amplitude damping errors. Regarding the complexity of our simulation, it should be noted that simulating the virtual distillation circuit for the two copy 6-qubit systems translates into a $2^{13}\times 2^{13}$ density matrix simulation, which is equivalent to a $26$ qubit pure state simulation.
%=============================================
% PART ONE
%=============================================
\subsection{Virtual distillation applied to variational states}
To approximate the ground state of $H_\mathrm{MaxCut}$ with QAOA, the variational state
\begin{equation}
    \rho(\boldsymbol{\alpha},\boldsymbol{\beta}) = U(\boldsymbol{\alpha},\boldsymbol{\beta})(\dyad{+})^{\otimes N}U^\dagger(\boldsymbol{\alpha},\boldsymbol{\beta})
\end{equation}
is prepared, where
\begin{equation}
    U(\boldsymbol{\alpha},\boldsymbol{\beta})=\prod_{j=1}^p e^{-i\beta_j H_M}e^{-i\alpha_j H_\mathrm{MaxCut}}
\end{equation}
is a unitary operation, $H_M = \sum_{i}X_i$ is a sum of Pauli $X$ matrices, $\ket{+}^{\otimes N}$ is a uniform superposition of all computational basis states, and $\alpha_j$, $\beta_j$ are $2p$ variational parameters. The variational parameters are optimized with respect to the expectation value of the MaxCut Hamiltonian
\begin{equation}
    \label{eq:costfunction}
    C(\boldsymbol{\alpha},\boldsymbol{\beta}):=\Tr(H_\mathrm{MaxCut}\rho(\boldsymbol{\alpha},\boldsymbol{\beta})),
\end{equation}
such that its value is minimized
\begin{equation}
    \label{eq:variational}
    (\boldsymbol{\alpha}_\mathrm{opt},\boldsymbol{\beta}_\mathrm{opt}) := \argmin_{\boldsymbol{\alpha},\boldsymbol{\beta}}C(\boldsymbol{\alpha},\boldsymbol{\beta}).
\end{equation}
We simulate a noisy QAOA circuit for $p=1$ where $e^{-i\alpha H_\mathrm{MaxCut}}$ is implemented as a product of \textsc{ZZ}--rotations, $\prod_{j,k\in E}e^{-i\alpha Z_jZ_k/2}$, and the mixer $e^{-i\beta H_M}$ is implemented as single qubit \textsc{X}--rotations, $\prod_{j} e^{-i\beta X_j}$. For the dephasing channel we have implemented the single-qubit error channel $\Lambda_\mathrm{dep}$ after the mixer gate to take into consideration the fact that dephasing errors do not commute with the X-rotation gate, resulting in an effect on the noise channel that will depolarize it and make it less noise biased, see \mbox{\figref[b]{fig:qaoa-circuit}}. In practical qubit platforms, two-qubit gates are generally more noisy than single-qubit gates~\cite{tannu_not_2019}. Thus, we reduce the error probability $\epsilon$ by a factor $10$ for the single-qubit gates. When simulating the QAOA with the amplitude damping channel $\Lambda_\mathrm{amp}$, we use it for both the one and two-qubit gates with $10\%$ less error probability for the single qubit gates. Finally, the same single-qubit error channel $\Lambda$ that is used in the QAOA circuit is also used in the virtual distillation circuit. We label the noisy expectation value for error channel obtained using QAOA without virtual distillation as $C^\Lambda(\alpha,\beta)$ and with virtual distillation as $C^\Lambda_\mathrm{mitigated}(\alpha,\beta)$. 
\par
We find the optimal variational parameters for both $C^\Lambda(\alpha,\beta)$ and $C^\Lambda_\mathrm{mitigated}(\alpha,\beta)$ for $21$ different error probabilities $\epsilon$ that are equally spaced between $0$ and $0.1$. To do this we start by optimizing the energy expectation value for no errors using brute-force optimization on a $100\times 100$ grid with $\alpha\in [0,\pi]$ and $\beta\in[0,\pi/2]$ together with an optimization/polishing function that uses the optimal grid point as an initial guess. The optimal parameters found for the noiseless QAOA circuit are then used as initial guess to the optimization function for the first noise iteration, and the initial guess is iteratively updated for each increasing noise level based on the optimal parameters found for the previous noise level. With this strategy, we aim to reduce the search space and computational time required for optimization at each subsequent noise level. Previous studies have indeed suggested that optimal parameters for QAOA circuits tend to remain relatively stable in the presence of moderate noise~\mbox{\cite{xue_effects_2019,sharma_noise_2020}}, providing another rationale for our iterative optimization approach.
%================BEGIN FIGURE=================
\begin{figure}[t]
    \centering
    \includegraphics[width=\linewidth]{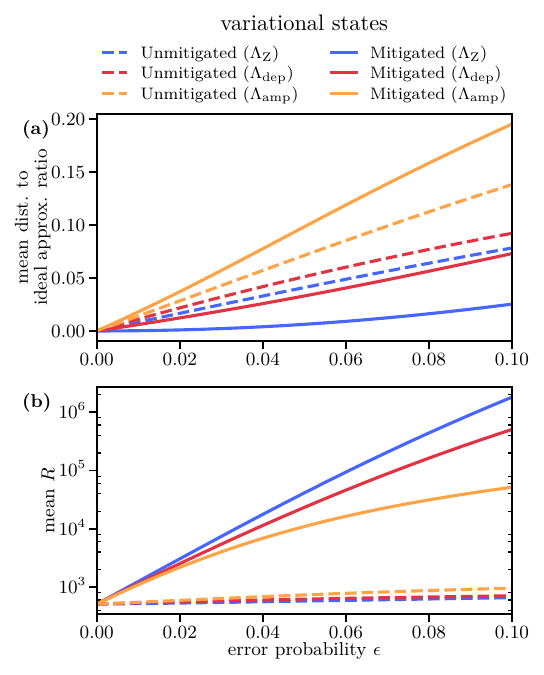}
    \caption{\textbf{(a)} The distance from the ideal approximation ratio with respect to the error probability $\epsilon$ averaged over all the instances. \textbf{(b)} The minimum number of samples $R$ required for the variance of the estimator to be $\leq 10^{-3}$ averaged over all the instances. Solid (dashed) lines are obtained with (without) virtual distillation. Blue lines are with respect to dephasing errors, red lines are with respect to depolarizing errors, and yellow lines with respect to amplitude damping.}
    \label{fig:qaoa}
\end{figure}
%=================END FIGURE==================
%================BEGIN FIGURE=================
\begin{figure}[t]
    \centering
    \includegraphics[width=\linewidth]{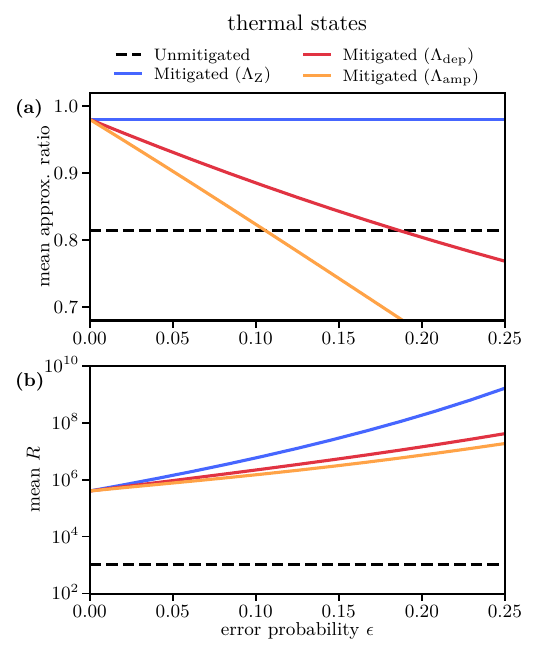}
    \caption{\textbf{(a)} The approximation ratio with respect to the error probability $\epsilon$ averaged over all the instances. \textbf{(b)} The minimum number of samples $R$ required for the variance of the estimator to be $\leq 10^{-3}$ averaged over all the instances. Solid (dashed) lines are obtained with (without) virtual distillation. Blue lines are with respect to dephasing errors, red lines are with respect to depolarizing errors, and yellow lines with respect to amplitude damping.}
    \label{fig:thermal}
\end{figure}
%=================END FIGURE==================
\par
After the optimization we compute the approximation ratio by dividing the expected cuts $C(\alpha_\mathrm{opt},\beta_\mathrm{opt})$, and $C^\Lambda_\mathrm{mitigated}(\alpha_\mathrm{opt},\beta_\mathrm{opt})$ by the maximum cut $C_\mathrm{max}$, where both expected cuts are calculated explicitly. \figref[a]{fig:qaoa} presents the difference between the ideal (noiseless) approximation ratio and the noisy approximation ratio averaged over the 30 instances without virtual distillation (unmitigated) for QAOA with the three types of error-channels, and with virtual distillation (mitigated) for the three types of error-channels. It can clearly be seen that the best error reduction is obtained for dephasing-errors in both the QAOA circuit and the virtual distillation circuit. From the figure the error reduction at $\epsilon=0.1$ with virtual distillation is $67\%$ when the underlying source of noise is dephasing errors, which shows excellent error suppression given a high error probability $\epsilon$. For depolarizing errors the error reduction is $20\%$ with virtual distillation. For amplitude damping errors we see no improvement, and actually the virtual distillation procedure degrades the results. This phenomenon presents an intriguing aspect of quantum error dynamics in the virtual distillation circuit that is not yet fully understood.
Interestingly, we find that dephasing-errors in the QAOA circuit yield a moderately better mean approximation ratio compared to depolarizing-errors without virtual distillation, as seen by the blue dashed line being below the red dashed line. This indicates that QAOA by itself may also be more robust against dephasing errors.
\par
It is known that single-qubit errors can lead to a coherent mismatch of the dominant eigenvector with respect to the ideal state~\cite{koczor_exponential_2021,huggins_virtual_2021,koczor_dominant_2021}. This means that the fidelity between dominant eigenvector of $\rho$ and the ideal noiseless state is not unity. In \appref{app:drift} we do a numerical analysis of the coherent mismatch caused by the errors in the QAOA circuit, and find that it is smaller for dephasing errors in the circuit. 
\par
In \figref[b]{fig:qaoa} we plot the minimum number of samples $R$ required for the variance of the estimator defined in Eq.~\eqref{eq:estimator} to be $\leq 10^{-3}$, averaged over all the 30 instances. We also do this for the unmitigated variance of the sample mean, given by
\begin{multline}
    \label{eq:unmitigated-variance}
    \Var(\mathrm{estim.}) = \frac{1}{2R}\Big[\Tr(\rho_\Lambda H_\mathrm{MaxCut}^2)
    \\ - \Tr(\rho_\Lambda H_\mathrm{MaxCut})^2\Big],
\end{multline}
where $R$ is the total number of samples and $\rho_\Lambda$ is the output from the QAOA circuit with error channel $\Lambda\in \{\Lambda_\mathrm{dep}, \Lambda_\mathrm{Z}\}$. Since the virtual distillation circuit uses two copies of $\rho_\Lambda$ as input but is only counted as one sample, therefore to provide an adjusted comparison that takes into account the additional resources used in virtual distillation, we scale the variance of the unmitigated sample mean by $1/2$. Indeed, for a pure state $\rho$, it can be shown that Eq.~\eqref{eq:mitigated-variance} is equal to Eq.~\eqref{eq:unmitigated-variance}~\cite{huggins_virtual_2021}. From the two solid lines in \figref[b]{fig:qaoa}, we see that the mean number of samples grow exponentially with virtual distillation for both types of errors. However, the difference between the number of samples needed for dephasing noise is only negligibly ($\ll$ factor of 10) larger than those needed for depolarizing noise.
%=============================================
% PART TWO
%=============================================
\subsection{Virtual distillation applied to therm\-al sta\-tes}
In the previous section, the density matrices that was given as input to the virtual distillation circuit were different as a result of the error-channel in the QAOA circuit. In this section, we make a noise agnostic comparison where the input state is the same to both virtual distillation circuits. The state that we have chosen is a statistical mixture between the state corresponding to the maximum-cut and a thermal state. The state corresponding the maximum-cut is given by the degenerate ground state $\rho_\mathrm{ideal}=\dyad{\psi_\mathrm{GS}}$ of the MaxCut Hamiltonian. We then consider the thermal state
\begin{equation}
    \rho_\mathrm{thermal} = \frac{e^{-\eta H_\mathrm{MaxCut}}}{Z(\eta)},
\end{equation}
where $Z(\eta)=\Tr(e^{-\eta H_\mathrm{MaxCut}})$ is the partition function, and $\eta$ is a constant that is proportional to the inverse temperature. From this we create an equally mixed state between the two states $\rho_\mathrm{ideal}$ and $\rho_\mathrm{thermal}$,
\begin{equation}
    \label{eq:thermal}
    \rho = \frac{1}{2}\rho_\mathrm{ideal} + \frac{1}{2}\rho_\mathrm{thermal}.
\end{equation}
Using $\rho$ as our input to the virtual distillation circuit we perform virtual distillation on it using the circuit in \figref[a]{fig:vd} with $\Lambda$ as either $\Lambda_\mathrm{Z}$, $\Lambda_\mathrm{dep}$ or $\Lambda_\mathrm{amp}$. In the simulations we choose $\eta=0.1$, and vary the noise-level $\epsilon$ between $0$ and $0.25$ and calculate the mitigated approximation ratio as well as the number of repetitions $R$ required for the variance of the estimator to be $\leq 10^{-3}$, averaged over all the 30 instances. \figref[a]{fig:thermal} shows the mean approximation ratio. While the mitigated approximation ratio stays constant for dephasing errors irrespective of the error probability $\epsilon$, the mitigated approximation ratio for both depolarizing errors and amplitude damping quickly decreases and even becomes lower than the unmitigated approximation ratio as seen by the solid red and yellow line crossing the black dashed line. This is because, while the MaxCut Hamiltonian commutes with the error operator $Z$ for the pure dephasing channel, it does not commute with all the error operators for the depolarizing and amplitude damping one. \figref[b]{fig:thermal} shows that the mean number of samples for the mitigated expectation values grows rapidly with the error probability. However, as before, the number of samples required when noise is pure dephasing is only slightly larger ($\lesssim$ factor of 10) than those required when noise is depolarizing. Note that the two solid lines do not intercept the dashed line at $\epsilon=0$, which is because the state $\rho$ in Eq.~\eqref{eq:thermal} is not a pure state.
%=============================================
% CONCLUSION
%=============================================
\section{\label{sec:conclusion}Conclusion}
In this work, we have studied the effects of dep\-olarizing-, dephasing-errors and amplitude damping in virtual distill\-ation circuits. We found that depolarizing errors in the circuit implementation of virtual distillation substantially degrade the mitigated expectation value. Additionally, it was observed that amplitude damping is extremely detrimental for virtual distillation resulting in no error mitigation when estimating the expectation value of the MaxCut Hamiltonian. For dephasing errors, we found that the quality of the mitigated expectation value does not degrade. This makes virtual distillation implemented in a system that strongly favors biased noise particular robust to errors. Moreover, trading biased noise for depolarizing errors does not significantly increase the sampling cost.
\par
Our findings suggest that implementing the virtual distillation protocol in a system that is strongly biased towards dephasing noise thus avoids the need to rely on other error mitigation techniques in addition to virtual distillation for mitigating errors. Finally, we mention that bias-preserving CSWAP gates can be implemented in bosonic cat code systems~\cite{puri_bias_2020,guillaud_repetition_2019} and Rydberg atoms~\cite{iris_hardware_2022}, which is crucial in order to not unbias or depolarize the noise channel.

%
% ****** Acknowledgments ******
\acknowledgments
We thank Oliver Hahn, Timo Hillman, Shahnawaz Ahmed, and Robert Jonsson for fruitful discussions. This work is supported from the Knut and Alice Wallenberg Foundation through the Wallenberg Center for Quantum Technology (WACQT). G. F. acknowledges support from the Swedish Research Council (Vetenskapsrådet) Grant QuACVA. SP was supported by the Air Force Office of Scientific Research under award number FA9550-21-1-0209.

%
% ****** Code availability ******
\codeavailability
\indent The code used for producing the results is made available in Ref.~\cite{pontus_2022}. All circuit simulations are done using Cirq~\cite{cirq_developers_2021} and Numpy~\cite{harris2020array}. As optimization function for the QAOA, we used \textsc{minimize} implemented in Scipy~\cite{virtanen_scipy_2020} with the BFGS algorithm as the default optimizer. The random graphs were generated using NetworkX~\cite{hagberg_exploring_2008}.

%
% ****** References ******
\bibliographystyle{quantum}
\bibliography{references}

\begin{thebibliography}{10}

\bibitem{lidar_quantum_2013}
Daniel~A Lidar and Todd~A Brun.
\newblock ``Quantum error correction''.
\newblock \href{https://dx.doi.org/10.1017/CBO9781139034807}{Cambridge University Press}. ~(2013).

\bibitem{li_efficient_2017}
Ying Li and Simon~C. Benjamin.
\newblock ``Efficient variational quantum simulator incorporating active error minimization''.
\newblock \href{https://dx.doi.org/10.1103/PhysRevX.7.021050}{Phys. Rev. X {\bf 7}, 021050}~(2017).

\bibitem{temme_error_2017}
Kristan Temme, Sergey Bravyi, and Jay~M. Gambetta.
\newblock ``Error mitigation for short-depth quantum circuits''.
\newblock \href{https://dx.doi.org/10.1103/PhysRevLett.119.180509}{Phys. Rev. Lett. {\bf 119}, 180509}~(2017).

\bibitem{endo2018practical}
Suguru Endo, Simon~C Benjamin, and Ying Li.
\newblock ``Practical quantum error mitigation for near-future applications''.
\newblock \href{https://dx.doi.org/10.1103/PhysRevX.8.031027}{Phys. Rev. X {\bf 8}, 031027}~(2018).

\bibitem{kandala2019error}
Abhinav Kandala, Kristan Temme, Antonio~D C{\'o}rcoles, Antonio Mezzacapo, Jerry~M Chow, and Jay~M Gambetta.
\newblock ``Error mitigation extends the computational reach of a noisy quantum processor''.
\newblock \href{https://dx.doi.org/10.1038/s41586-019-1040-7}{Nature {\bf 567}, 491--495}~(2019).

\bibitem{strikis2021learning}
Armands Strikis, Dayue Qin, Yanzhu Chen, Simon~C Benjamin, and Ying Li.
\newblock ``Learning-based quantum error mitigation''.
\newblock \href{https://dx.doi.org/10.1103/PRXQuantum.2.040330}{PRX Quantum {\bf 2}, 040330}~(2021).

\bibitem{czarnik2021error}
Piotr Czarnik, Andrew Arrasmith, Patrick~J Coles, and Lukasz Cincio.
\newblock ``Error mitigation with clifford quantum-circuit data''.
\newblock \href{https://dx.doi.org/10.22331/q-2021-11-26-592}{Quantum {\bf 5}, 592}~(2021).

\bibitem{bonet2018low}
Xavi Bonet-Monroig, Ramiro Sagastizabal, M~Singh, and T.~E. O'Brien.
\newblock ``Low-cost error mitigation by symmetry verification''.
\newblock \href{https://dx.doi.org/10.1103/PhysRevA.98.062339}{Phys. Rev. A {\bf 98}, 062339}~(2018).

\bibitem{mcardle2019error}
Sam McArdle, Xiao Yuan, and Simon Benjamin.
\newblock ``Error-mitigated digital quantum simulation''.
\newblock \href{https://dx.doi.org/10.1103/PhysRevLett.122.180501}{Phys. Rev. Lett. {\bf 122}, 180501}~(2019).

\bibitem{sagastizabal2019experimental}
R.~Sagastizabal, X.~Bonet-Monroig, M.~Singh, M.~A. Rol, C.~C. Bultink, X.~Fu, C.~H. Price, V.~P. Ostroukh, N.~Muthusubramanian, A.~Bruno, M.~Beekman, N.~Haider, T.~E. O'Brien, and L.~DiCarlo.
\newblock ``Experimental error mitigation via symmetry verification in a variational quantum eigensolver''.
\newblock \href{https://dx.doi.org/10.1103/PhysRevA.100.010302}{Phys. Rev. A {\bf 100}, 010302(R)}~(2019).

\bibitem{google2020hartree}
{Google AI Quantum and Collaborators}, Frank Arute, Kunal Arya, Ryan Babbush, Dave Bacon, Joseph~C Bardin, Rami Barends, Sergio Boixo, Michael Broughton, Bob~B Buckley, et~al.
\newblock ``Hartree-fock on a superconducting qubit quantum computer''.
\newblock \href{https://dx.doi.org/10.1126/science.abb9811}{Science {\bf 369}, 1084--1089}~(2020).

\bibitem{huggins2021efficient}
William~J Huggins, Jarrod~R McClean, Nicholas~C Rubin, Zhang Jiang, Nathan Wiebe, K~Birgitta Whaley, and Ryan Babbush.
\newblock ``Efficient and noise resilient measurements for quantum chemistry on near-term quantum computers''.
\newblock \href{https://dx.doi.org/10.1038/s41534-020-00341-7}{npj Quantum Inf {\bf 7}, 1--9}~(2021).

\bibitem{o2021error}
Thomas~E O’Brien, Stefano Polla, Nicholas~C Rubin, William~J Huggins, Sam McArdle, Sergio Boixo, Jarrod~R McClean, and Ryan Babbush.
\newblock ``Error mitigation via verified phase estimation''.
\newblock \href{https://dx.doi.org/10.1103/PRXQuantum.2.020317}{PRX Quantum {\bf 2}, 020317}~(2021).

\bibitem{maciejewski2020mitigation}
Filip~B Maciejewski, Zolt{\'a}n Zimbor{\'a}s, and Micha{\l} Oszmaniec.
\newblock ``Mitigation of readout noise in near-term quantum devices by classical post-processing based on detector tomography''.
\newblock \href{https://dx.doi.org/10.22331/q-2020-04-24-257}{Quantum {\bf 4}, 257}~(2020).

\bibitem{cotler_quantum_2019}
Jordan Cotler, Soonwon Choi, Alexander Lukin, Hrant Gharibyan, Tarun Grover, M.~Eric Tai, Matthew Rispoli, Robert Schittko, Philipp~M. Preiss, Adam~M. Kaufman, Markus Greiner, Hannes Pichler, and Patrick Hayden.
\newblock ``Quantum virtual cooling''.
\newblock \href{https://dx.doi.org/10.1103/PhysRevX.9.031013}{Phys. Rev. X {\bf 9}, 031013}~(2019).

\bibitem{koczor_exponential_2021}
Bálint Koczor.
\newblock ``Exponential error suppression for near-term quantum devices''.
\newblock \href{https://dx.doi.org/10.1103/PhysRevX.11.031057}{Phys. Rev. X {\bf 11}, 031057}~(2021).

\bibitem{huggins_virtual_2021}
William~J. Huggins, Sam McArdle, Thomas~E. O'Brien, Joonho Lee, Nicholas~C. Rubin, Sergio Boixo, K.~Birgitta Whaley, Ryan Babbush, and Jarrod~R. McClean.
\newblock ``Virtual distillation for quantum error mitigation''.
\newblock \href{https://dx.doi.org/10.1103/PhysRevX.11.041036}{Phys. Rev. X {\bf 11}, 041036}~(2021).

\bibitem{dumitrescu_cloud_2018}
E.~F. Dumitrescu, A.~J. {McCaskey}, G.~Hagen, G.~R. Jansen, T.~D. Morris, T.~Papenbrock, R.~C. Pooser, D.~J. Dean, and P.~Lougovski.
\newblock ``Cloud quantum computing of an atomic nucleus''.
\newblock \href{https://dx.doi.org/10.1103/PhysRevLett.120.210501}{Phys. Rev. Lett. {\bf 120}, 210501}~(2018).

\bibitem{otten_recovering_2019}
Matthew Otten and Stephen~K. Gray.
\newblock ``Recovering noise-free quantum observables''.
\newblock \href{https://dx.doi.org/10.1103/PhysRevA.99.012338}{Phys. Rev. A {\bf 99}, 012338}~(2019).

\bibitem{giurgica-tiron_digital_2020}
Tudor Giurgica-Tiron, Yousef Hindy, Ryan {LaRose}, Andrea Mari, and William~J. Zeng.
\newblock ``Digital zero noise extrapolation for quantum error mitigation''.
\newblock \href{https://dx.doi.org/10.1109/QCE49297.2020.00045}{2020 {IEEE} International Conference on Quantum Computing and Engineering ({QCE})Pages 306--316}~(2020).

\bibitem{jnane_multicore_2022}
Hamza Jnane, Brennan Undseth, Zhenyu Cai, Simon~C. Benjamin, and B\'alint Koczor.
\newblock ``Multicore quantum computing''.
\newblock \href{https://dx.doi.org/10.1103/PhysRevApplied.18.044064}{Phys. Rev. Appl. {\bf 18}, 044064}~(2022).

\bibitem{puri_bias_2020}
Shruti Puri, Lucas St-Jean, Jonathan~A. Gross, Alexander Grimm, Nicholas~E. Frattini, Pavithran~S. Iyer, Anirudh Krishna, Steven Touzard, Liang Jiang, Alexandre Blais, Steven~T. Flammia, and S.~M. Girvin.
\newblock ``Bias-preserving gates with stabilized cat qubits''.
\newblock \href{https://dx.doi.org/10.1126/sciadv.aay5901}{Sci Adv {\bf 6}, eaay5901}~(2020).

\bibitem{grimm2020stabilization}
Alexander Grimm, Nicholas~E Frattini, Shruti Puri, Shantanu~O Mundhada, Steven Touzard, Mazyar Mirrahimi, Steven~M Girvin, Shyam Shankar, and Michel~H Devoret.
\newblock ``Stabilization and operation of a kerr-cat qubit''.
\newblock \href{https://dx.doi.org/10.1038/s41586-020-2587-z}{Nature {\bf 584}, 205--209}~(2020).

\bibitem{iris_hardware_2022}
Iris Cong, Harry Levine, Alexander Keesling, Dolev Bluvstein, Sheng-Tao Wang, and Mikhail~D. Lukin.
\newblock ``Hardware-efficient, fault-tolerant quantum computation with rydberg atoms''.
\newblock \href{https://dx.doi.org/10.1103/PhysRevX.12.021049}{Phys. Rev. X {\bf 12}, 021049}~(2022).

\bibitem{smolin_1996_five}
John~A. Smolin and David~P. DiVincenzo.
\newblock ``Five two-bit quantum gates are sufficient to implement the quantum fredkin gate''.
\newblock \href{https://dx.doi.org/10.1103/PhysRevA.53.2855}{Phys. Rev. A {\bf 53}, 2855}~(1996).

\bibitem{koczor_dominant_2021}
B{\'a}lint Koczor.
\newblock ``The dominant eigenvector of a noisy quantum state''.
\newblock \href{https://dx.doi.org/10.1088/1367-2630/ac37ae}{New J. Phys. {\bf 23}, 123047}~(2021).

\bibitem{aharonov_polynomial_2009}
Dorit Aharonov, Vaughan Jones, and Zeph Landau.
\newblock ``A polynomial quantum algorithm for approximating the jones polynomial''.
\newblock \href{https://dx.doi.org/10.1007/s00453-008-9168-0}{Algorithmica {\bf 55}, 395--421}~(2009).

\bibitem{nielsen_quantum_2010}
Michael~A. Nielsen and Isaac~L. Chuang.
\newblock ``Quantum computation and quantum information: 10th anniversary edition''.
\newblock \href{https://dx.doi.org/10.1017/CBO9780511976667}{Cambridge University Press}. ~(2010).

\bibitem{farhi_qaoa_2014}
Edward Farhi, Jeffrey Goldstone, and Sam Gutmann.
\newblock ``A quantum approximate optimization algorithm''~(2014).
\newblock  \href{http://arxiv.org/abs/1411.4028}{arXiv:1411.4028}.

\bibitem{lucas_ising_2014}
Andrew Lucas.
\newblock ``Ising formulations of many {NP} problems''.
\newblock \href{https://dx.doi.org/10.3389/fphy.2014.00005}{Front. Physics {\bf 2}, 5}~(2014).

\bibitem{vikstal_2020}
Pontus Vikst\aa{}l, Mattias Gr\"onkvist, Marika Svensson, Martin Andersson, G\"oran Johansson, and Giulia Ferrini.
\newblock ``Applying the quantum approximate optimization algorithm to the tail-assignment problem''.
\newblock \href{https://dx.doi.org/10.1103/PhysRevApplied.14.034009}{Phys. Rev. Applied {\bf 14}, 034009}~(2020).

\bibitem{xia_electronic}
Rongxin Xia, Teng Bian, and Sabre Kais.
\newblock ``Electronic structure calculations and the ising hamiltonian''.
\newblock \href{https://dx.doi.org/10.1021/acs.jpcb.7b10371}{J. Phys. Chem. B {\bf 122}, 3384--3395}~(2018).

\bibitem{czarnik_qubit-efficient_2021}
Piotr Czarnik, Andrew Arrasmith, Lukasz Cincio, and Patrick~J. Coles.
\newblock ``Qubit-efficient exponential suppression of errors''~(2021).
\newblock  \href{http://arxiv.org/abs/2102.06056}{arXiv:2102.06056}.

\bibitem{erdos_random_1959}
P.~Erdős and A.~Rényi.
\newblock ``On random graphs. i.''.
\newblock \href{https://dx.doi.org/10.5486/PMD.1959.6.3-4.12}{Publ. Math. Debrecen {\bf 6}, 290--297}~(1959).

\bibitem{tannu_not_2019}
Swamit~S. Tannu and Moinuddin~K. Qureshi.
\newblock ``Not all qubits are created equal''.
\newblock In Proceedings of the Twenty-Fourth International Conference on Architectural Support for Programming Languages and Operating Systems.
\newblock \href{https://dx.doi.org/10.1145/3297858.3304007}{Pages 987--999}.
\newblock {ACM}~(2019).

\bibitem{xue_effects_2019}
Cheng Xue, Zhao-Yun Chen, Yu-Chun Wu, and Guo-Ping Guo.
\newblock ``Effects of quantum noise on quantum approximate optimization algorithm''~(2019).
\newblock  \href{http://arxiv.org/abs/1909.02196}{arxiv:1909.02196}.

\bibitem{sharma_noise_2020}
Kunal Sharma, Sumeet Khatri, M~Cerezo, and Patrick~J Coles.
\newblock ``Noise resilience of variational quantum compiling''.
\newblock \href{https://dx.doi.org/10.1088/1367-2630/ab784c}{New J. Phys. {\bf 22}, 043006}~(2020).

\bibitem{guillaud_repetition_2019}
Jérémie Guillaud and Mazyar Mirrahimi.
\newblock ``Repetition cat qubits for fault-tolerant quantum computation''.
\newblock \href{https://dx.doi.org/10.1103/PhysRevX.9.041053}{Phys. Rev. X {\bf 9}, 041053}~(2019).

\bibitem{pontus_2022}
Pontus Wikståhl.
\newblock ``Wikstahl/error-mitigation: Version 2.0 (v2.0)''.
\newblock {Z}enodo, \url{https://doi.org/10.5281/zenodo.7215577}~(2022).

\bibitem{cirq_developers_2021}
{Cirq Developers}.
\newblock ``Cirq (v0.12.0)''.
\newblock {Z}enodo, \url{https://doi.org/10.5281/zenodo.5182845}~(2021).

\bibitem{harris2020array}
Charles~R. Harris, K.~Jarrod Millman, St{\'{e}}fan~J. van~der Walt, Ralf Gommers, Pauli Virtanen, David Cournapeau, Eric Wieser, Julian Taylor, Sebastian Berg, Nathaniel~J. Smith, Robert Kern, Matti Picus, Stephan Hoyer, Marten~H. van Kerkwijk, Matthew Brett, Allan Haldane, Jaime~Fern{\'{a}}ndez del R{\'{i}}o, Mark Wiebe, Pearu Peterson, Pierre G{\'{e}}rard-Marchant, Kevin Sheppard, Tyler Reddy, Warren Weckesser, Hameer Abbasi, Christoph Gohlke, and Travis~E. Oliphant.
\newblock ``Array programming with {NumPy}''.
\newblock \href{https://dx.doi.org/10.1038/s41586-020-2649-2}{Nature {\bf 585}, 357--362}~(2020).

\bibitem{virtanen_scipy_2020}
Pauli Virtanen, Ralf Gommers, Travis~E. Oliphant, Matt aberland, Tyler Reddy, David Cournapeau, Evgeni Burovski, Pearu Peterson, Warren Weckesser, Jonathan Bright, St{\'e}fan~J. {van der Walt}, Matthew Brett, Joshua Wilson, K.~Jarrod Millman, Nikolay Mayorov, Andrew R.~J. Nelson, Eric Jones, Robert Kern, Eric Larson, C~J Carey, {\.I}lhan Polat, Yu~Feng, Eric~W. Moore, Jake {VanderPlas}, Denis Laxalde, Josef Perktold, Robert Cimrman, Ian Henriksen, E.~A. Quintero, Charles~R. Harris, Anne~M. Archibald, Ant{\^o}nio~H. Ribeiro, Fabian Pedregosa, Paul {van Mulbregt}, and {SciPy 1.0 Contributors}.
\newblock ``{{SciPy} 1.0: Fundamental Algorithms for Scientific Computing in Python}''.
\newblock \href{https://dx.doi.org/10.1038/s41592-019-0686-2}{Nature Methods {\bf 17}, 261--272}~(2020).

\bibitem{hagberg_exploring_2008}
Aric~A. Hagberg, Daniel~A. Schult, and Pieter~J. Swart.
\newblock ``Exploring network structure, dynamics, and function using networkx''.
\newblock Presented at the 7th Python in Science Conference, Pasadena, CA USA~(2008).
\newblock pp. 11-15.

\bibitem{kempen_mean_2000}
G.M.P. van Kempen and L.J. van Vliet.
\newblock ``Mean and variance of ratio estimators used in fluorescence ratio imaging''.
\newblock \href{https://dx.doi.org/10.1002/(SICI)1097-0320(20000401)39:4<300::AID-CYTO8>3.0.CO;2-O}{Cytometry {\bf 39}, 300--305}~(2000).

\end{thebibliography}

%
% ****** Supplementary Material ******
\onecolumn
\newpage
\appendix

%=============================================
% MITIGATED EXPECTATION VALUE
%=============================================
\section{\label{app:mitigated-expvals}Error mitigated expectation value}
In this section we derive an analytical expression of how the mitigated expectation value is affected by single-qubit depolarizing and dephasing errors that occur during the virtual distillation circuit. To simplify the derivation it is useful to use the following convention for the single-qubit error channels:
\begin{align}
    \Lambda_\mathrm{dep}(\rho) &= \qty(1-\frac{3}{4}\epsilon)\rho + \frac{3}{4}\epsilon\qty(X\rho X + Y\rho Y+Z\rho Z), \\
    \Lambda_\mathrm{Z}(\rho) &= \qty(1-\frac{\epsilon}{2})\rho + \frac{\epsilon}{2}Z\rho Z,
\end{align}
which amounts to substituting $\epsilon\rightarrow 4\epsilon/3$ in the end to get back to Eq.~\eqref{eq:depolarizing-channel} for the depolarizing channel, and $\epsilon\rightarrow 2\epsilon$ to get back to Eq.~\eqref{eq:dephasing-channel} for the dephasing channel.
\par
We start by deriving the noisy mitigated expectation value for two-copies virtual distillation and then generalize the results to an even number of copies. The output from the virtual distillation circuit in \figref[a]{fig:vd} can be written as
\begin{equation}
    \rho_\mathrm{out} 
    = \underbrace{\tilde \Lambda_N\circ \mathcal{U}_N\circ \tilde \Lambda_{N-1}\circ \mathcal{U}_{N-1}\ldots \tilde \Lambda_1\circ \mathcal{U}_1}_{\Lambda_\mathrm{tot}}\rho_\mathrm{in} 
    = \Lambda_\mathrm{tot}(\rho_\mathrm{in}),
\end{equation}
where $\rho_\mathrm{in}$ is the input state
\begin{equation}
    \rho_\mathrm{in} := \dyad{+}\otimes\rho\otimes\rho,
\end{equation}
$\mathcal{U}_i(\rho_\mathrm{in}) = \mathrm{CSWAP}_{i}\, \rho_\mathrm{in}\, \mathrm{CSWAP}_{i}^\dagger$ is the controlled-$\mathrm{SWAP}$ gate that swaps the $i$th qubit of the two subsystems $\rho^{\otimes 2}$ conditioned on the auxiliary qubit being in the state $\ket{1}$, and is given by
\begin{equation}
     \mathrm{CSWAP}_{i} = \dyad{0}\otimes I^{\otimes N}\otimes I^{\otimes N} + \dyad{1}\otimes\mathrm{SWAP}_{i},
\end{equation}
where $I$ is a $2\times 2$ identity matrix; $\tilde \Lambda_i$ is the product of the single-qubit error channels that are applied to the qubits involved in the gate $\mathcal{U}_i$,
\begin{equation}
    \tilde \Lambda_i = \Lambda \otimes \Lambda_i \otimes \Lambda_i = \Lambda \otimes \Lambda^{\otimes 2}_i,
\end{equation}
with $\Lambda\in\{\Lambda_\mathrm{dep},\Lambda_\mathrm{Z}\}$ being a single-qubit error channel, and the subscript $i$ of $\Lambda_i$ indicates that $\Lambda$ acts on the $i$th qubit of subsystem $\rho$. Using the cyclic permutation of the trace the expectation value of the numerator in Eq.~\eqref{eq:estimated} for $M=2$ can be expressed as
\begin{equation}
    \label{eq:numerator}
    \Tr(X_\mathrm{aux} O^{(2)}\Lambda_\mathrm{tot}(\rho_\mathrm{in}))
    = \Tr(\Lambda_\mathrm{tot}^\dagger(X_\mathrm{aux} O^{(2)})\rho_\mathrm{in}),
\end{equation}
where
\begin{equation}
    \label{eq:inverse-channel}
    \Lambda^\dagger_\mathrm{tot} := \mathcal{U}^\dagger_1\circ \tilde\Lambda^\dagger_1 \circ \mathcal{U}^\dagger_{2}\circ \tilde\Lambda^\dagger_{2} \ldots \mathcal{U}^\dagger_N \circ \tilde\Lambda^\dagger_N
    = \circ_{k=1}^N (\mathcal{U}^\dagger_k\circ\tilde\Lambda^\dagger_k)
\end{equation}
is the adjoint of $\Lambda_\mathrm{tot}$, $X_\mathrm{aux}$ is the Pauli $X$ operator on the auxiliary qubit, and $O^{(2)}$ acts on the two subsystems $\rho$. It can be noted that Eq.~\eqref{eq:numerator} is reminiscent of the Heisenberg picture in which the operators are evolving instead of the quantum states. We can furthermore drop $\dagger$ from $\mathcal{U}_k$ and $\Lambda_k$ since they are both Hermitian quantum maps.
\par
Beginning with the action of $\tilde \Lambda_N$ in Eq.~\eqref{eq:inverse-channel} on $X_\mathrm{aux}O^{(2)}$, the result is
\begin{align}
    \label{eq:auxo}
    \tilde \Lambda_N(X_\mathrm{aux} O^{(2)}) = \Lambda(X) \otimes \Lambda_N^{\otimes 2}(O^{(2)})
    = (1-\epsilon) X\otimes \Lambda_N^{\otimes 2}(O^{(2)}),
\end{align}
where we have used the fact that 
\begin{equation}
    \Lambda(X) = \Lambda(\dyad{0}{1}) + \Lambda(\dyad{1}{0})
    = (1-\epsilon)\dyad{0}{1} + (1-\epsilon)\dyad{1}{0} 
    = (1-\epsilon)X,
\end{equation}
for $\Lambda\in\{\Lambda_\mathrm{dep}, \Lambda_\mathrm{Z}\}$. Next, Eq.~\eqref{eq:auxo} is followed by the action of $\mathcal{U}_N$ which gives
\begin{equation}
    (\mathcal{U}_N\circ \tilde\Lambda_N)(X_\mathrm{aux} O^{(2)}) = (1-\epsilon)\qty[\dyad{1}{0}\otimes \mathrm{SWAP}_N \Lambda_N^{\otimes 2}(O^{(2)}) + \dyad{0}{1}\otimes \Lambda_N^{\otimes 2}(O^{(2)})\mathrm{SWAP}_N].
\end{equation}
Repeating this for the remaining terms yields
\begin{align}
    \Lambda^\dagger_\mathrm{tot}(X_\mathrm{aux} O^{(2)}) &= (1-\epsilon)^{N}\qty[\dyad{1}{0} \otimes\prod_{k=1}^N \mathrm{SWAP}_k \bar \Lambda^{\otimes 2}(O^{(2)})
    + \dyad{0}{1} \otimes \bar \Lambda^{\otimes 2}(O^{(2)}) \prod_{k=1}^N \mathrm{SWAP}_k] \notag \\
    \label{eq:channel}
    &= (1-\epsilon)^N X\otimes S^{(2)}\bar \Lambda^{\otimes 2}(O^{(2)}),
\end{align}
where $\bar\Lambda^{\otimes 2}:=\bar \Lambda \otimes \bar \Lambda = (\Lambda_1\circ\ldots\circ\Lambda_N)\otimes (\Lambda_1\circ\ldots\circ\Lambda_N) = \Lambda\otimes\ldots\otimes\Lambda$ is a tensor product of $2N$ single-qubit error channels $\Lambda$, and where we have used the fact that $S^{(2)}=\prod_{k=1}^N \mathrm{SWAP}_k$ and $[S^{(2)},\bar \Lambda^{\otimes 2}(O^{(2)})]=0$ in the last equality. The expectation value of the numerator is thus calculated to be
\begin{align}
    \Tr(\Lambda^\dagger_\mathrm{tot}(X_\mathrm{aux} O^{(2)}) \rho_\mathrm{in})
    = (1-\epsilon)^N\Tr(X\dyad{+})\Tr(\bar\Lambda^{\otimes 2}(O^{(2)})S^{(2)} \rho^{\otimes 2})
    = (1-\epsilon)^N \Tr(\bar\Lambda(O)\rho^2).
\end{align}
To calculate the denominator of Eq.~\eqref{eq:estimated} for $M=2$, we simply replace $O$ with the identity $I^{\otimes N}$ for which $\bar\Lambda(I^{\otimes N})=I^{\otimes N}$ because both the depolarizing- and dephasing-channel are unital maps. Hence the mitigated expectation value with either depolarizing or dephasing-errors is given by
\begin{equation}
    \label{eq:mitigated-expectation-value-two-copies}
    \expval{O}^\Lambda_\mathrm{mitigated}
    = \frac{\Tr(\Lambda^\dagger_\mathrm{tot}(X_\mathrm{aux} O^{(2)}) \rho_\mathrm{in})}{\Tr(\Lambda^\dagger_\mathrm{tot}(X_\mathrm{aux}) \rho_\mathrm{in})}
    = \frac{(1-\epsilon)^N}{(1-\epsilon)^N} \frac{\Tr(\bar\Lambda(O)\rho^2)}{\Tr(\rho^2)}
    = \frac{\Tr(\bar\Lambda(O)\rho^2)}{\Tr(\rho^2)}.
\end{equation}
We have reached the final expression which shows that single-qubit depolarizing or dephasing errors that occur during the circuit are equivalent to a tensor product of $N$-single qubit depolarizing or dephasing channels $\Lambda$ acting on the observable $O$. 
\par
It is useful to consider what happens for the special case when $O$ is a tensor product of Pauli-operators, e.g. $O=X\otimes Z\otimes I\otimes Y\ldots $. In this case we have that $\Lambda_\mathrm{dep}(\sigma) = \qty(1-\epsilon)\sigma$, with $\sigma\in\{X,Y,Z\}$, for the depolarizing channel, and $\Lambda_\mathrm{Z}(\sigma) = \qty(1-\epsilon)\sigma$, with $\sigma\in\{X,Y\}$, for the dephasing channel. Hence the mitigated expectation value for the two channels can be written as
\begin{align}
    \expval{O}^{\Lambda_\mathrm{dep}}_\mathrm{mitigated} &= (1-\epsilon)^k\frac{\Tr(O\rho^2)}{\Tr(\rho^2)},
    \\
    \expval{O}^{\Lambda_\mathrm{Z}}_\mathrm{mitigated} &= (1-\epsilon)^{k'}\frac{\Tr(O\rho^2)}{\Tr(\rho^2)},
\end{align}
where $k$ is the number of $\{X,Y,Z\}$ Pauli-operators in the tensor product of $O$ and $k'$ is the number of $\{X,Y\}$ Pauli-operators in the tensor product of $O$. With the substitution of $\epsilon$ back to the definitions of Eq.~\eqref{eq:depolarizing-channel} and Eq.~\eqref{eq:dephasing-channel} the mitigated expectation values becomes
\begin{align}
    \expval{O}^{\Lambda_\mathrm{dep}}_\mathrm{mitigated} &= \qty(1-\frac{4}{3}\epsilon)^k\frac{\Tr(O\rho^2)}{\Tr(\rho^2)},
    \\
    \expval{O}^{\Lambda_\mathrm{Z}}_\mathrm{mitigated} &= (1-2\epsilon)^{k'}\frac{\Tr(O\rho^2)}{\Tr(\rho^2)}.
\end{align}
These results can easily be generalized for arbitrary even $M$. This is because for even $M$ one can execute controlled-swaps such that every subsystem $\rho$ is only involved in one controlled-swap operations, which is not the case for odd $M$. For instance, with four copies of $\rho$, one can execute controlled-swaps between subsystems one and two, and three and four, respectively, as depicted in \mbox{\figref[b]{fig:vd}}. The importance is to ensure that no copy remains in its original position, thus achieving the necessary derangement. Consequently, the auxiliary qubit is engaged in $NM/2$ controlled-swaps. Therefore the expression for the mitigated expectation value for even $M$ becomes
\begin{align}
    \expval{O}^\Lambda_\mathrm{mitigated}
    = \frac{\Tr(\Lambda^\dagger_\mathrm{tot}(X_\mathrm{aux} O^{(M)}) \rho_\mathrm{in})}{\Tr(\Lambda^\dagger_\mathrm{tot}(X_\mathrm{aux}) \rho_\mathrm{in})}
    = \frac{(1-\epsilon)^{NM/2}}{(1-\epsilon)^{NM/2}} \frac{\Tr(\bar\Lambda(O)\rho^M)}{\Tr(\rho^M)}
    = \frac{\Tr(\bar\Lambda(O)\rho^M)}{\Tr(\rho^M)},
\end{align}
which is the same expression as Eq.~\eqref{eq:mitigated-expectation-value-two-copies}, but for arbitrary even $M$.
%=============================================
% VARIANCE OF THE ESTIMATOR
%=============================================
\section{\label{app:variance}Variance of the estimator with dephasing errors}
In this section, we derive the variance of the estimator in Eq.~\eqref{eq:estimator} with dephasing errors in the virtual distillation circuit for $M=2$. While there exists no closed-form expression for the variance of the quotient between two random variables, an approximate expression can be obtained by doing a Taylor expansion around the mean of the two random variables~\cite{kempen_mean_2000}. By doing so, one finds that the variance of the estimator is
\begin{equation}
    \Var(\mathrm{estim.})\approx 
    \frac{1}{R}\qty(\frac{\Var(X_\mathrm{aux} O^{(2)})}{\expval{X_\mathrm{aux}}^2} + \frac{\expval{X_\mathrm{aux} O^{(2)}}^2}{\expval{X_\mathrm{aux}}^4}\Var(X_\mathrm{aux}) - 2\frac{\expval{O^{(2)}X_\mathrm{aux}}}{\expval{X_\mathrm{aux}}^3}\Cov(O^{(2)}X_\mathrm{aux},X_\mathrm{aux})),
\end{equation}
where $R$ is the number of samples. The terms $\expval{X_\mathrm{aux}}$ and $\expval{O^{(2)}X_\mathrm{aux}}$ have been evaluated in \appref{app:mitigated-expvals}, and we are thus left to evaluate the three following terms
\begin{align}
    \Var(X_\mathrm{aux}) &= \expval*{(X_\mathrm{aux})^2} - \expval*{X_\mathrm{aux}}^2 = 1 - \qty(1-2\epsilon)^{2N}\Tr(\rho^2)^2, 
    \\
    \Cov(O^{(2)}X_\mathrm{aux},X_\mathrm{aux}) &= \expval*{O^{(2)}X_\mathrm{aux}^2} - \expval*{O^{(2)}X_\mathrm{aux}}\expval*{X_\mathrm{aux}} \notag \\
    &= \expval*{O^{(2)}} - (1-2\epsilon)^{2N}\Tr(\bar\Lambda_\mathrm{Z}(O)\rho^2)\Tr(\rho^2) \notag \\
    &= \Tr(\bar\Lambda_\mathrm{Z}(O)\rho) - (1-2\epsilon)^{2N}\Tr(\bar\Lambda_\mathrm{Z}(O)\rho^2)\Tr(\rho^2).
    \\
    \Var(O^{(2)}X_\mathrm{aux}) &= \expval*{(O^{(2)}X_\mathrm{aux})^2} - \expval*{O^{(2)}X_\mathrm{aux}}^2 \notag 
    \\
    \label{eq:varox}
    &= \expval*{(O^{(2)})^2} - \qty(1-2\epsilon)^{2N}\Tr(\bar\Lambda_\mathrm{Z}(O)\rho^2)^2 \notag \\
    &= \frac{1}{4}\expval{(O^\mathbf{1})^2}+\frac{1}{4}\expval{(O^\mathbf{2})^2}+\frac{1}{2}\expval{O^\mathbf{1}O^\mathbf{2}} - \qty(1-2\epsilon)^{2N}\Tr(\bar\Lambda_\mathrm{Z}(O)\rho^2)^2 \notag \\
    &= \frac{1}{2}\Tr(\bar\Lambda_\mathrm{Z}(O^2) \rho) + \frac{1}{2}\Tr(\bar\Lambda_\mathrm{Z}(O)\rho)^2 - \qty(1-2\epsilon)^{2N}\Tr(\bar\Lambda_\mathrm{Z}(O)\rho^2)^2,
\end{align}
where in the last step of Eq.~\eqref{eq:varox} we have used the following fact to evaluate $\expval{O^\mathbf{1}O^\mathbf{2}} = \Tr(\Lambda_\mathrm{tot}^\dagger(O^\mathbf{1}O^\mathbf{2})\rho_\mathrm{in})$
\begin{align}
    \Lambda_\mathrm{tot}^\dagger(O^\mathbf{1}O^\mathbf{2}) &= \circ_{k=1}^N(\mathcal{U}^\dagger_k \circ \tilde \Lambda_k^\dagger)(O^\mathbf{1}O^\mathbf{2}) \notag
    \\
    &= \dyad{0}\otimes \bar\Lambda_\mathrm{Z}(O)^{\otimes 2}
    + \dyad{1}\otimes \qty(\prod_{k=1}^N\mathrm{SWAP}_k)\bar\Lambda_\mathrm{Z}(O)^{\otimes 2}\qty(\prod_{k=1}^N\mathrm{SWAP}_k) \notag
    \\
    &= \dyad{0}\otimes \bar\Lambda_\mathrm{Z}(O)^{\otimes 2}
    + \dyad{1}\otimes \bar\Lambda_\mathrm{Z}(O)^{\otimes 2} \notag
    \\
    &= I \otimes \bar\Lambda_\mathrm{Z}(O)^{\otimes 2},
\end{align}
and hence $\Tr(\Lambda_\mathrm{tot}^\dagger(O^\mathbf{1}O^\mathbf{2})\rho_\mathrm{in})=\Tr(\dyad{+})\Tr(\bar\Lambda_\mathrm{Z}(O)\rho)^2 =\Tr(\bar\Lambda_\mathrm{Z}(O)\rho)^2$. Given these calculations we get that the variance of the estimator for dephasing errors in the virtual distillation circuit is
\begin{multline}
    \label{eq:variance-estimator}
    \Var(\mathrm{estim.})=
    \frac{1}{R(1-2\epsilon)^{2N}}
    \Bigg(\frac{\Tr(\bar\Lambda_\mathrm{Z}(O^2)\rho)}{2\Tr(\rho^2)^2}
    + \frac{\Tr(\bar\Lambda_\mathrm{Z}(O)\rho)^2}{2\Tr(\rho^2)^2}
    + \frac{\Tr(\bar\Lambda_\mathrm{Z}(O)\rho^2)^2}{\Tr(\rho^2)^4}
    \\
    - 2\frac{\Tr(\bar\Lambda_\mathrm{Z}(O)\rho^2)}{\Tr(\rho^2)^3}\Tr(\bar\Lambda_\mathrm{Z}(O)\rho)\Bigg).
\end{multline}
Comparing this equation with the expression for the noiseless variance of the estimator given by Eq.~\eqref{eq:mitigated-variance}, we see that Eq.~\eqref{eq:variance-estimator} is scaled by a factor $1/(1-2\epsilon)^{2N}$ and that $O$ has been replaced by $\bar \Lambda_\mathrm{Z}(O)$.
%=============================================
% COHERENT MISMATCH
%=============================================
\section{\label{app:drift}Coherent mismatch}
In this section we investigate the coherent mismatch, also known as drift, of the dominant eigenvector that is caused by the noise in the QAOA circuit. Since virtual distillation relies on the assumption that the dominant eigenvector approximates the ideal state, it is of importance to quantify how good of an approximation this is. The coherent mismatch is defined as~\cite{koczor_dominant_2021}
\begin{equation}
    c := 1 - \abs*{\braket{\psi_\mathrm{dom}}{\psi_\mathrm{ideal}}}^2,
\end{equation}
where $\ket{\psi_\mathrm{dom}}$ is the dominant eigenvector of a noisy quantum state $\rho$ and $\ket{\psi_\mathrm{ideal}}$ is with respect to the ideal (noiseless) state. In \figref[a]{fig:drift} we plot the coherent mismatch, averaged over all instances, between the ideal noiseless QAOA state, and the dominant eigenvector of $\rho_\Lambda(\alpha_\mathrm{opt},\beta_\mathrm{opt})$, where the optimal angles are with respect to $C_\mathrm{mitigated}^\Lambda(\alpha,\beta)$. From the figure it can be seen that the coherent mismatch is smaller for dephasing errors in the QAOA circuit compared to both depolarizing errors and amplitude damping. For the case of depolarizing errors this fact can be explained by converting an incoherent error to a coherent one. Consider a $Z$ error that occurs before the $X(\beta)$-gate. By commuting the error through the gate we get
\begin{equation}
    X(\beta)Z = e^{-i\beta X}Z = \cos(\beta)Z-i\sin(\beta)XZ
    = ZX(-\beta),
\end{equation}
where in the last equality we have used the anti-commutation relation between $X$ and $Z$, $XZ=-ZX$. Hence, we see that a $Z$ error introduces an incoherent error that corresponds to an extra $\pi$ rotation of the $X(\beta)$-gate. For the $ZZ(\alpha)$-gate a $Z$ error acting on any of the two-qubits commute through the gate, and hence only the $X(\beta)$-gate will be affected by dephasing errors which causes coherent mismatch. For the depolarizing channel, however, both $X$ and $Y$ errors will lead to a coherent error of the $ZZ(\alpha)$-gate which will result in a greater coherent mismatch. Nevertheless, we see that the coherent mismatch is small for large error probabilities, less than $2\%$ for depolarizing error.
%================BEGIN FIGURE=================
\begin{figure}[ht]
    \centering
    \includegraphics{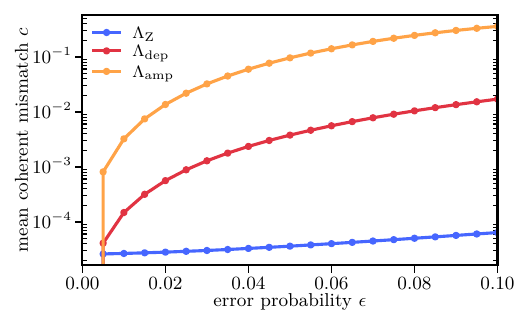}
    \caption{The coherent mismatch averaged over all instances. The blue line is the mean coherent mismatch with dephasing errors in the QAOA circuit and the red line is with depolarizing errors in the QAOA circuit.}
    \label{fig:drift}
\end{figure}
%=================END FIGURE==================
\end{document}